\documentclass[]{spie}  

\usepackage{amsmath,amsfonts,amssymb}
\usepackage{graphicx}
\usepackage[colorlinks=true, allcolors=blue]{hyperref}

\title{The Photonic Lantern Nuller: from concept to laboratory and on-sky demonstrations}

\author[a,b]{Yinzi Xin}
\author[a]{Nemanja Jovanovic}
\author[a]{Dimitri Mawet}
\author[a]{Daniel Echeverri}
\author[c]{Yoo Jung Kim}
\author[d]{Jonathan Lin}
\author[e]{Julien Lozi}
\author[e,f]{Sébastien Vievard}
\author[e]{Olivier Guyon}
\author[g]{Grace Piroscia}
\author[e]{Vincent Déo}
\author[g]{Sergio Leon-Saval}
\author[h]{Rodrigo Amezcua-Correa}
\author[h]{Stephanos Yerolatsitis}
\author[b]{Sebastiaan Y. Haffert}
\author[c]{Michael P. Fitzgerald}
\author[c]{Pradip Gatkine}
\author[i]{Suvinay Goyal}
\author[h]{Barnaby Norris}
\author[j]{Garreth Ruane}
\author[k]{Steph Sallum}
\affil[a]{Department of Astronomy, Caltech, 1200 E. California Blvd, Pasadena, USA}
\affil[b]{Sterrewacht Leiden, PO Box 9513, Niels Bohrweg 2, Leiden, The Netherlands}
\affil[c]{Department of Physics \& Astronomy, 430 Portola Plaza, University of California, Los Angeles, USA}
\affil[d]{NASA Ames Research Center, Moffett Field, USA}
\affil[e]{Subaru Telescope, National Astronomical Observatory of Japan, 650 N. Aohoku Place, Hilo, USA}
\affil[f]{Space Science and Engineering Initiative, College of Engineering, University of Hawai'i, 640 N. Aohoku Place, Hilo, USA}
\affil[g]{Sydney Institute for Astronomy, The University of Sydney, School of Physics A28, Sydney, Australia}
\affil[h]{The College of Optics and Photonics, University of Central Florida, 4304 Scorpius Street, Orlando, USA}
\affil[i]{Department of Astronomy, University of Illinois Urbana-Champaign, Urbana, USA}
\affil[j]{Jet Propulsion Laboratory, California Institute of Technology, 4800 Oak Grove Drive, Pasadena, USA}
\affil[k]{Department of Astronomy \& Astrophysics, 1156 High Street, University of California, Santa Cruz, USA}

\authorinfo{Further author information: (Send correspondence to Y.X.)\\Y.X.: E-mail: xin@strw.leidenuniv.nl}

\begin{document} 
\maketitle

\begin{abstract}
This thesis work presents the conceptual design and experimental characterization of the Photonic Lantern Nuller instrument, which uses a multimode-to-single-mode demultiplexing waveguide to cancel out starlight while maintaining planet light, allowing for the direct characterization of planets at a telescope’s diffraction limit. The PLN was experimentally characterized in the lab, where it was further enhanced using common-path wavefront sensing and control techniques, and then demonstrated on sky at the Subaru Telescope. Highlights include measured in-lab null-depths of $\sim10^{-4}$ in three out of four ports simultaneously and on-sky null-depths of $\sim10^{-1}$ (limited by jitter and atmospheric residuals). We provide an overview of these results and discuss avenues for future work.
\end{abstract}

\keywords{photonic lanterns, exoplanets, coronagraphy, nulling, astrophotonics}

\section{INTRODUCTION}
\label{sec:intro}  

High-resolution spectroscopy plays a central role in the effort to characterize exoplanets, enabling measurements of planetary radial velocities, rotational velocities, atmospheric composition, and surface inhomogeneities through Doppler imaging techniques \cite{wang_hdc1}. By measuring the radial velocities of planets, it may also provide a pathway toward the detection of exomoons \cite{ruffio_exomoons}. The Photonic Lantern Nuller (PLN) \cite{xin_2022, Tuthill2022-NIH} is a high-contrast nulling instrument designed to detect and characterize companions at angular separations approaching and extending within $1,\lambda/D$, where $\lambda$ is the observing wavelength and $D$ is the telescope diameter. The PLN is conceptually related to the Vortex Fiber Nuller (VFN) \cite{Ruane2018_VFN,echeverri_2019}. However, whereas the VFN provides a single nulled channel with an azimuthally symmetric response, the PLN generates four distinct nulled channels, each with a different spatial coupling pattern. This multiplexed response increases the overall companion throughput and provides additional information for constraining both the companion flux ratio and its location on the sky, subject to an inherent $180^{\circ}$ positional ambiguity \cite{xin_2022}. The same spatial diversity also enables spectroastrometric measurements \cite{kim_2024, kim_2025_onsky}, although with a mode-selective lantern, the corresponding astrometric solutions retain this degeneracy \cite{kim_2024}.

The PLN is implemented using a six-port mode-selective photonic lantern (MSPL) \cite{LeonSaval_MSPL}, a form of photonic lantern \cite{LeonSaval_PL_2013} that employs dissimilar core geometries to map individual inputs onto linearly polarized (LP) modes, corresponding to the eigenmodes of a weakly guiding, radially symmetric step-index waveguide. Each supported LP mode at the multimode input face of the lantern is adiabatically transformed by the lantern into a separate, dedicated single-mode fiber output. The spatial symmetries of several of these modes naturally produce zero coupling on-axis, allowing a star aligned to the lantern center to be rejected, while maintaining nonzero transmission for off-axis sources, such as planetary companions.

A detailed theoretical treatment of the PLN is presented in Ref.~\citenum{xin_2022}, while the first laboratory demonstration, including both monochromatic and broadband operation, is reported in Ref.~\citenum{xin_2024_pln_lab}. Additional laboratory work presented in Ref.~\citenum{xin_implicit_2025} used wavefront control methods to further improve the PLN null-depths by approximately two orders of magnitude. In this proceedings, we present preliminary results from an onsky demonstration of the PLN, which was installed on the Subaru Coronagraphic Extreme Adaptive Optics Instrument (SCExAO) instrument at Subaru Observatory in September 2024.

\section{DAYTIME CALIBRATION AND TESTING} \label{sec:daycal}
\begin{figure*}[t]
\begin{center}
	\includegraphics[scale = 0.6]{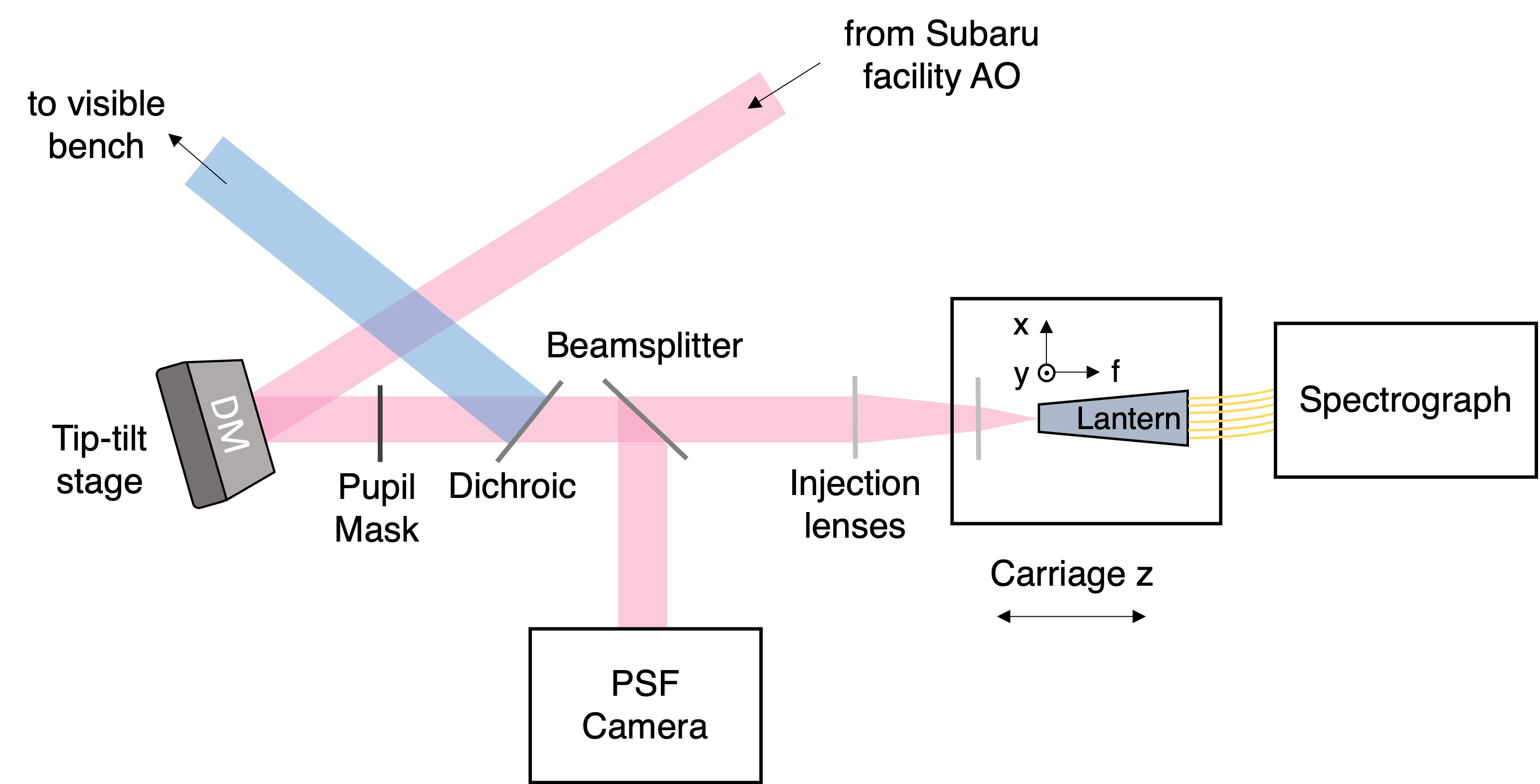}
	\caption{\label{fig:scexao_schematic} A simplified schematic of SCEXAO's optical path, including only the elements relevant for the near-infrared photonic lanterns. The photonics injection unit consists of a two-lens injection system (in which the carriage position for one of the lenses can be adjusted to control the $F\#$ of the beam), and stage to which lanterns or fibers can be mounted, which can also be translated in XY (in the plane perpendicular to the beam) as well as F (to control the lateral position relative to the focus of the beam). The outputs of the lantern are dispersed onto a spectrograph with a resolving power of approximately 160.}
\end{center}
\end{figure*}

The relevant submodules of the SCEXAO instrument are described and shown in the schematic in Fig. \ref{fig:scexao_schematic}. For a detailed description of the instrument, including submodules not discussed here, see Ref.~\citenum{ahn_scexao_2021}. SCExAO sits after AO3k, the primary adaptive optics (AO) system of the Subaru Telescope, and has a 2000-actuator deformable mirror and wavefront sensor that constitute a second-stage AO loop. For this work, the near-infrared light is sent to the photonics injection unit as well as a PSF monitoring camera in a 90:10 split. The lantern is mounted within a two-lens injection system that provides control over the $F\#$ of the beam. The lantern mount can also be translated in 2-dimensions in the plane perpendicular to the beam, as well as lateral to the beam (to adjust the focus). The outputs of the lantern are dispersed onto a spectrograph with a resolving power of approximately 160. The stage holding the DM can be used to adjust the alignment of the beam going into the infrared arm.

\begin{figure*}[t]
\begin{center}
	\includegraphics[scale = 0.67]{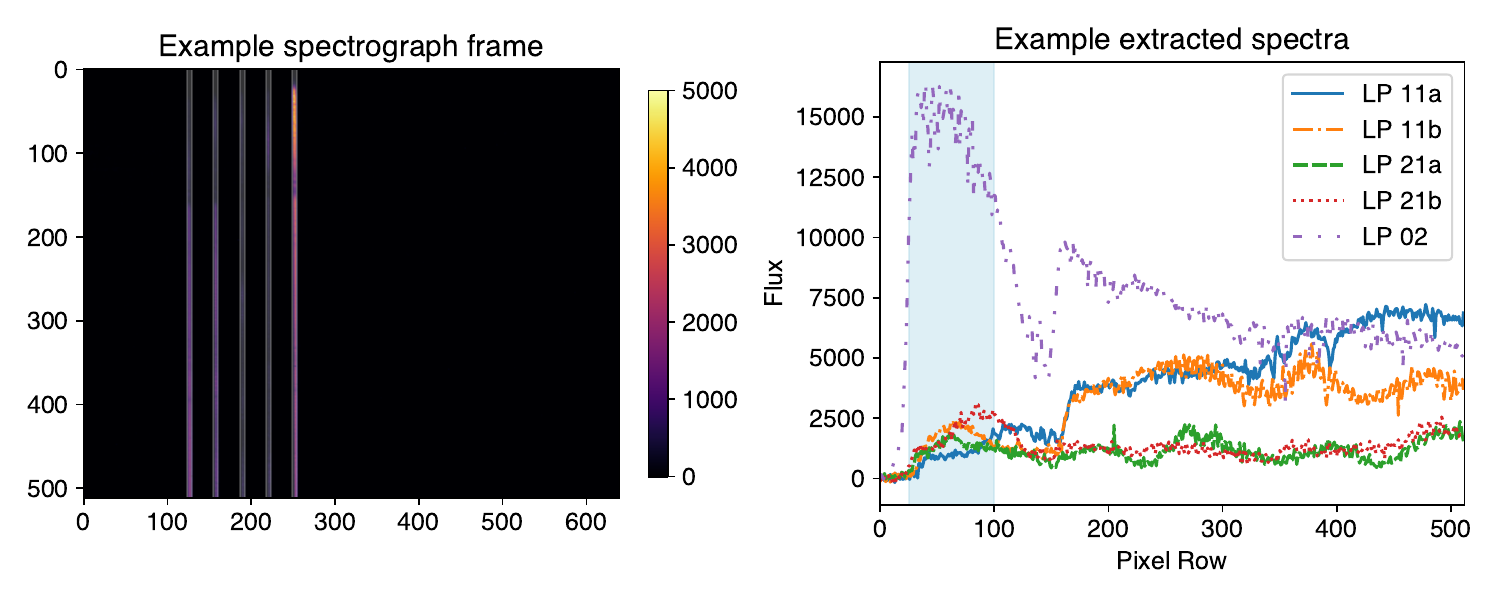}
	\caption{\label{fig:example_spectra} Left) An example frame obtained from the detector, showing the spectra from (from left to right) the LP 11a, LP 11b, LP 21a, LP 21b, and LP 02 ports. The shaded regions indicate the traces used for spectral extraction. Right) The extracted spectra from the traces indicated on the left, obtained by summing the counts in each row. The blue shading indicates the wavelength region summed over for computing coupling maps. Smaller row numbers correspond to longer wavelengths.}
\end{center}
\end{figure*}

An example data frame from the spectrograph is shown in Fig. \ref{fig:example_spectra}a, along with the traces used for spectral extraction. Example spectra for the five ports, extracted by summing across horizontally across each defined trace, are shown in Fig. \ref{fig:example_spectra}b. The resolving power of this spectrograph is $R\approx160$. Because the LP 01 port is very bright when the lantern is centered, its trace is placed off the detector to prevent saturation. However, the detector stage can be translated to measure the LP 01 port spectrum on its own.

During daytime testing, we calibrated the lantern to the beam. Optimizing the injection $F\#$ to maximize the peak coupling into the LP 11a port results in an $F\#$ of about 6, consistent with previous laboratory experiments presented in Refs. \citenum{xin_laboratory_2024, xin_implicit_2025}. An example frame on the spectrograph detector is shown in Fig. \ref{fig:example_spectra}a, along with the traces used for spectral extraction. Example spectra for the five ports, extracted by summing across each row, are shown in Fig. \ref{fig:example_spectra}b. 

\begin{figure*}[t]
\begin{center}
	\includegraphics[scale = 0.7]{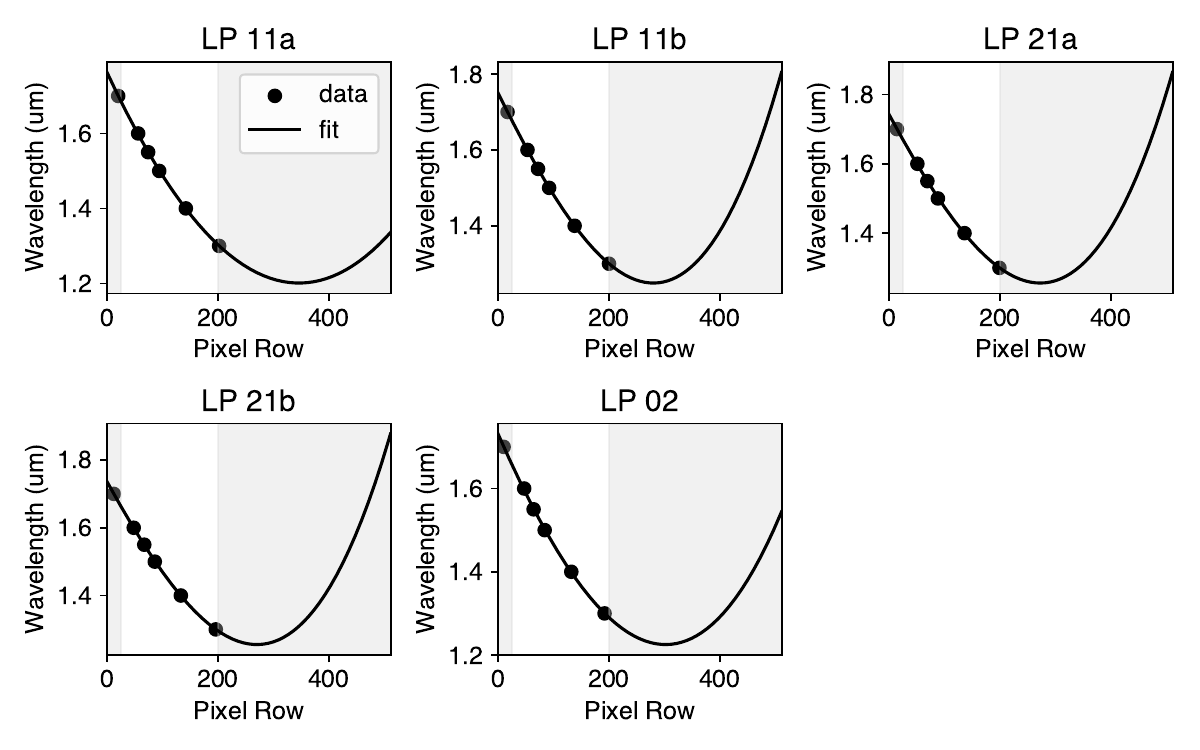}
	\caption{\label{fig:wavel_calib} Mapping of detector pixel row to wavelength, obtained by fitting a cubic polynomial to the location of peak intensities for each port. The calibration is valid between the 25th and 200th rows, or between approximately 1.30 $\mu$m and 1.65 $\mu$m. Thus, data from gray shaded regions are excluded from the rest of this work.}
\end{center}
\end{figure*}

To calibrate the wavelengths corresponding to the pixels, we insert a series of narrowband filters in front of the light source and measure the location of the peak intensity on the detector. The wavelength as a function of pixel row for each port is shown in Fig. \ref{fig:wavel_calib}, along with the best fit cubic polynomials. The overall transmission is very low below the 25th row, and there is an inversion in the fitted polynomial past the 200th row. The rest of this work thus only uses data between the 25th and 200th rows, or between approximately 1.30 $\mu$m and 1.65 $\mu$m.

The photonic lantern was optimized for operation at a wavelength of 1.55~$\mu$m. Coupling maps within the relevant spectral range were generated by integrating the flux recorded in rows 25 through 100 of each output port, corresponding approximately to wavelengths between 1.47~$\mu$m and 1.67~$\mu$m. Relative motion between the incident beam and the lantern can be achieved either by translating the injection stage or by introducing tip-tilt to the incoming wavefront. The latter produces a measurable displacement on the PSF monitoring camera, whose plate scale is calibrated in milliarcseconds, whereas the injection stage coordinates are not directly referenced to physical angular units. However, the available tip-tilt range is limited to roughly 50 mas about the optical axis and is therefore cannot sample the full coupling map. To establish a conversion between stage motion and angular displacement, a two-step calibration procedure was adopted.

First, the PSF was held fixed at the center of the monitoring camera while the injection stage was scanned in both orthogonal directions. The stage was then positioned at the location corresponding to the minimum total coupling across the four nulled ports. With the injection stage fixed at this null position, the PSF was scanned over a $100 \times 100$ mas field using tip-tilt control. An additional scan centered on the null was subsequently acquired by translating the injection stage. We fit the relative magnification and rotation between the resulting datasets, resulting in a conversion between stage displacement and angular units on the sky.

\begin{figure*}[t]
\begin{center}
	\includegraphics[scale = 0.5]{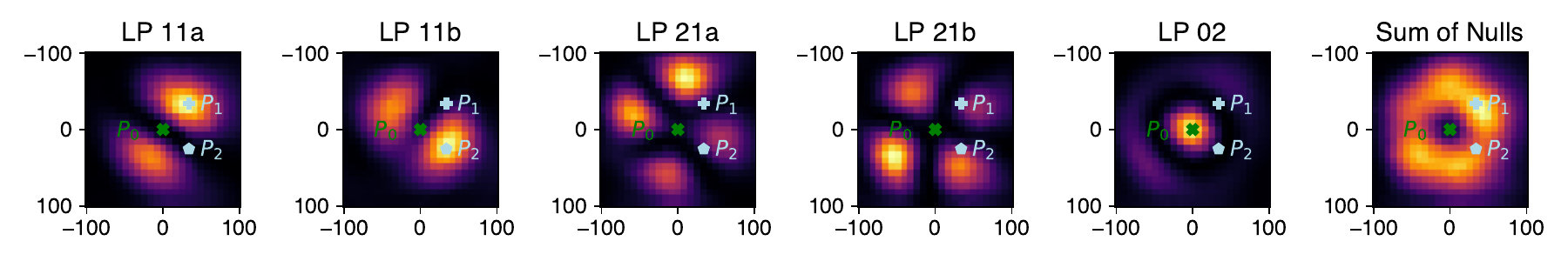}
    \includegraphics[scale = 0.5]{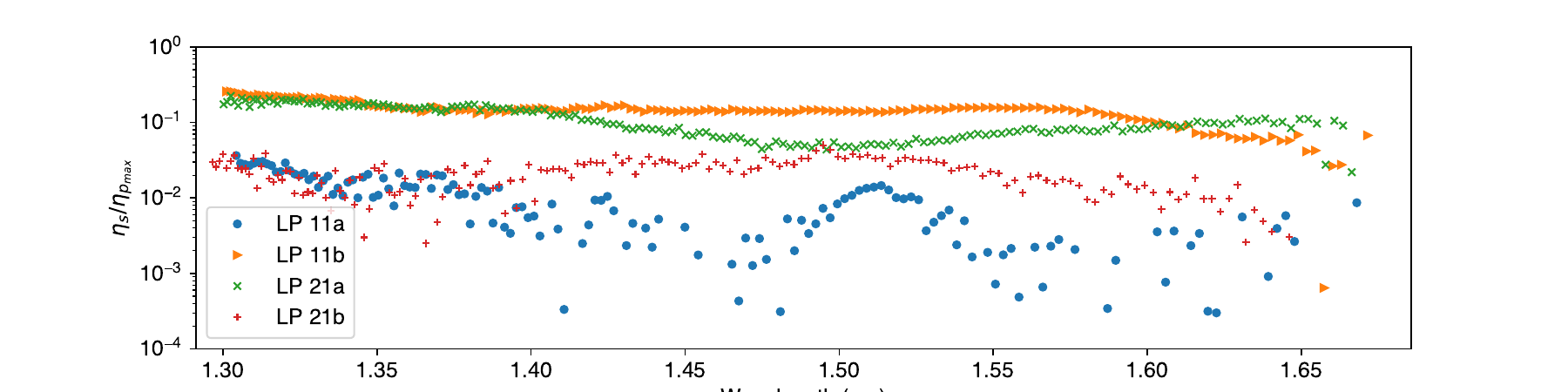}
	\caption{\label{fig:lab_maps} Top) Laboratory-measured coupling maps for the five output ports, shown on a spatial scale calibrated in mas. The far-right panel displays the total coupling obtained by summing the four nulled ports. Three reference locations are marked to indicate positions subsequently targeted during on-sky observations. Point $P_0$ marks the lantern center, corresponding to the minimum summed coupling and the location used to measure the stellar coupling efficiency. Points $P_1$ and $P_2$ denote the positions of maximum off-axis coupling for the LP11a and LP11b ports, respectively. Bottom) Wavelength-dependent null depths for each output port, computed as the ratio of stellar coupling to the corresponding peak coupling.}
\end{center}
\end{figure*}

The resulting coupling maps, obtained from lantern translations (and converted to milliarcsecond coordinates using this calibration), are presented in the top of Fig.~\ref{fig:lab_maps}. Three locations of interest are indicated, corresponding to positions later used during on-sky observations (Section~\ref{sec:onsky}). Point $P_0$ denotes the lantern center, where the target star is aligned and the stellar coupling efficiency, $\eta_s$, is measured. Points $P_1$ and $P_2$ identify the locations of maximum off-axis coupling, $\eta_{p_{\mathrm{peak}}}$, for the LP11a and LP11b outputs, respectively.

The wavelength-dependent null depths, defined as $\eta_s/\eta_{p_{peak}}$, are presented in the bottom of Fig.~\ref{fig:lab_maps} for each output port. For a given wavelength, $\eta_{p_{peak}}$ is evaluated at the spatial location corresponding to the maximum coupling of that port across the entire null bandwidth, as determined from the coupling maps shown in Fig.~\ref{fig:lab_maps} --- i.e. the peak-coupling position is held fixed across wavelength for each port rather than being independently optimized at each spectral channel.

Since SCExAO engineering nights involve the use of multiple instrument submodules and frequently changing hardware configurations, the calibrated lantern alignment cannot be maintained between observing sessions. Instead, the alignment settings are recorded and subsequently restored prior to on-sky observations. In practice, however, accurately recovering the alignment on-sky remains challenging. As discussed in Section~\ref{sec:onsky}, additional work is required to identify the sources of alignment offsets between daytime and on-sky operation and to develop more reliable procedures for recovering the optimal lantern alignment during observations.

\section{ON-SKY ENGINEERING} \label{sec:onsky}

\subsection{Misalignment and Attempted Recovery}

\begin{figure*}[t]
\begin{center}
    \includegraphics[scale = 0.57]{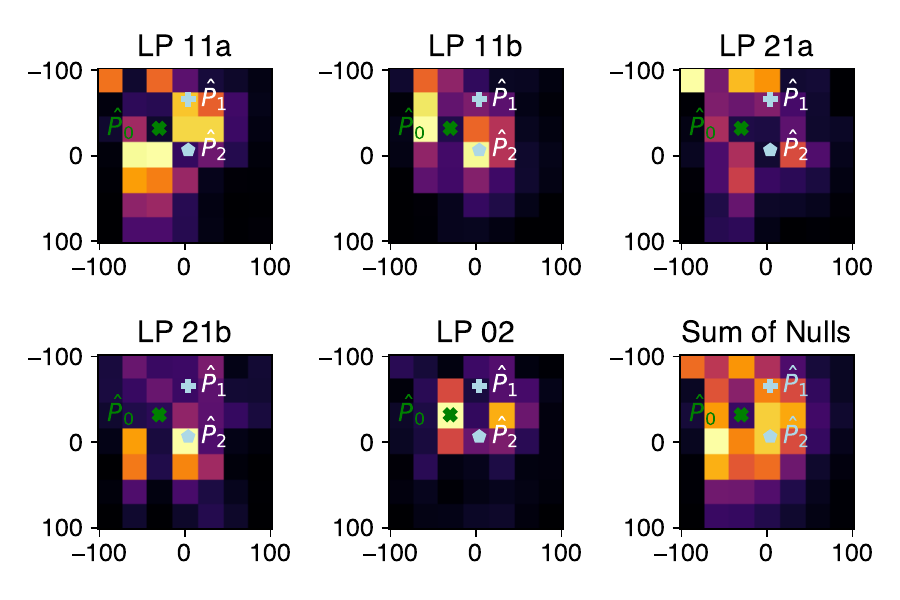}
    \includegraphics[scale = 0.47]{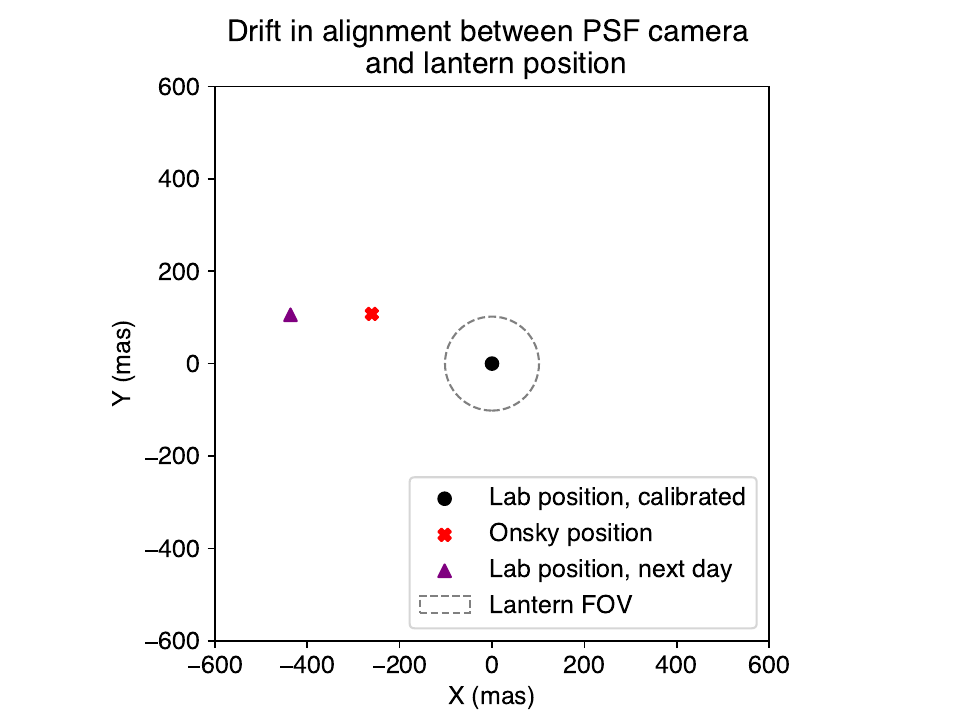}
	\caption{\label{fig:sky_alignment} Left) On-sky coupling maps for the five measured output ports, with spatial scale in mas. Longer measurements were subsequently acquired at the locations $\hat{P}_0$, $\hat{P}_1$, and $\hat{P}_2$, the on-sky estimates of the corresponding reference points defined in Fig.~\ref{fig:lab_maps}. Right) Measured displacement between the lantern position and the central pixel of the PSF monitoring camera. The nominal daytime calibration position is located at the origin and marked by the black circle. The gray circle indicates the approximate extent of the coupling-map scans shown in this figure and in Fig.~\ref{fig:lab_maps}, covering the lantern's field of view. The lantern position recovered during the on-sky observations is shown by the blue square, while the position measured on the following afternoon is marked by the green triangle. The observed offsets demonstrate a substantial alignment shift between calibration and observation, with the subsequent measurement indicating a continued drift in the same direction after observing.
}
\end{center}
\end{figure*}

During an engineering night on 11 March 2025, we observed HD144206, a B-type star with an H-band magnitude of 4.923. Initial measurements showed that the total flux coupled into the lantern was greater with the adaptive optics (AO) loop open than with it closed, indicating a substantial alignment error. To recover the alignment, the lantern position was adjusted using manual gradient-ascent until the stellar PSF was brought back within the lantern field of view (FOV). A subsequent scan produced the coupling maps shown in the left panel of Fig.~\ref{fig:sky_alignment}. The resulting maps exhibited a well-defined central minimum in the summed flux across the nulled ports. The lantern was aligned to this location, denoted $\hat{P}_0$, which serves as the on-sky estimate of the nominal center position $P_0$. Corresponding estimates of the off-axis coupling maxima are denoted by $\hat{P}_1$ and $\hat{P}_2$.

The offset in the injection unit location between the daytime calibration position and the recovered on-sky alignment is shown in the right panel of Fig.~\ref{fig:sky_alignment}, with injection-stage coordinates converted to mas. The initial alignment error was sufficiently large that the stellar beam fell entirely outside the lantern FOV. Measurements from the afternoon following the observations further suggest that the alignment continued to drift in the same direction after the engineering night. This drift is possibly due to thermal changes in the telescope and optical bench environment between daytime calibration and night time observing.

\subsection{On-sky null-depths}
\begin{figure*}[t]
\begin{center}
    \includegraphics[scale = 0.65]{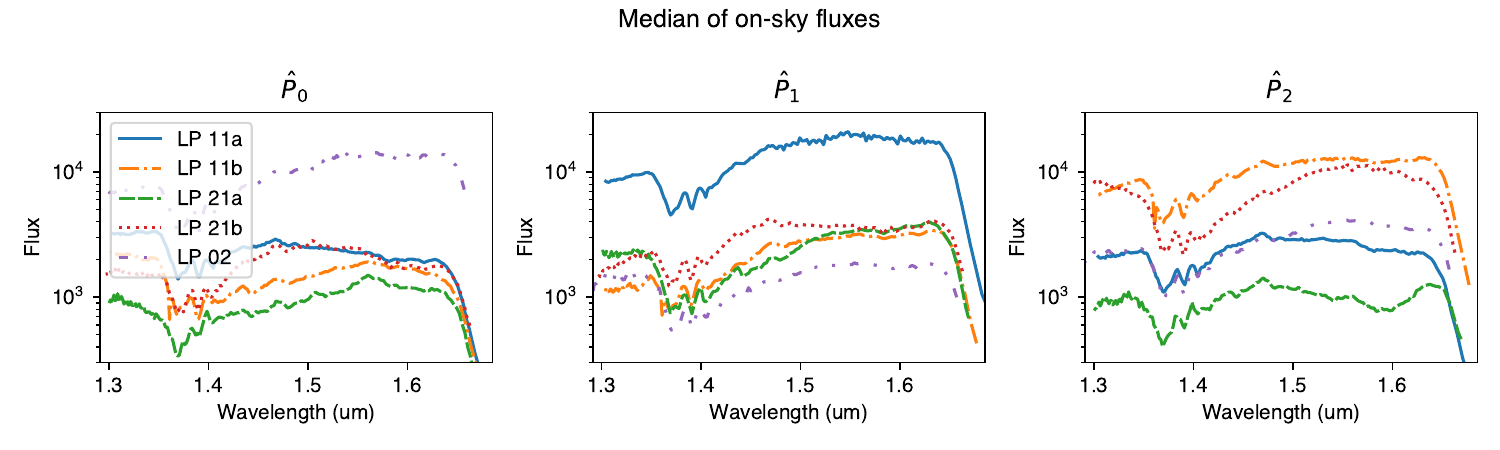}
	\caption{\label{fig:sky_data} The median coupled spectra at $\hat{P}_0$, $\hat{P}_1$, $\hat{P}_2$. At $\hat{P}_0$, port with the highest coupling is LP 02, as expected. Meanwhile, the LP 11a and LP 11b ports have the highest coupling at At $\hat{P}_1$ and At $\hat{P}_2$, also as expected.}
\end{center}
\end{figure*}

After approximately recovering the lantern alignment, we obtained further measurements of the coupling at the three points $\hat{P}_0$, $\hat{P}_1$, and $\hat{P}_2$. The medians of the resulting spectra are presented in Fig. \ref{fig:sky_data}.

\begin{figure*}[t]
\begin{center}
	\includegraphics[scale = 0.6]{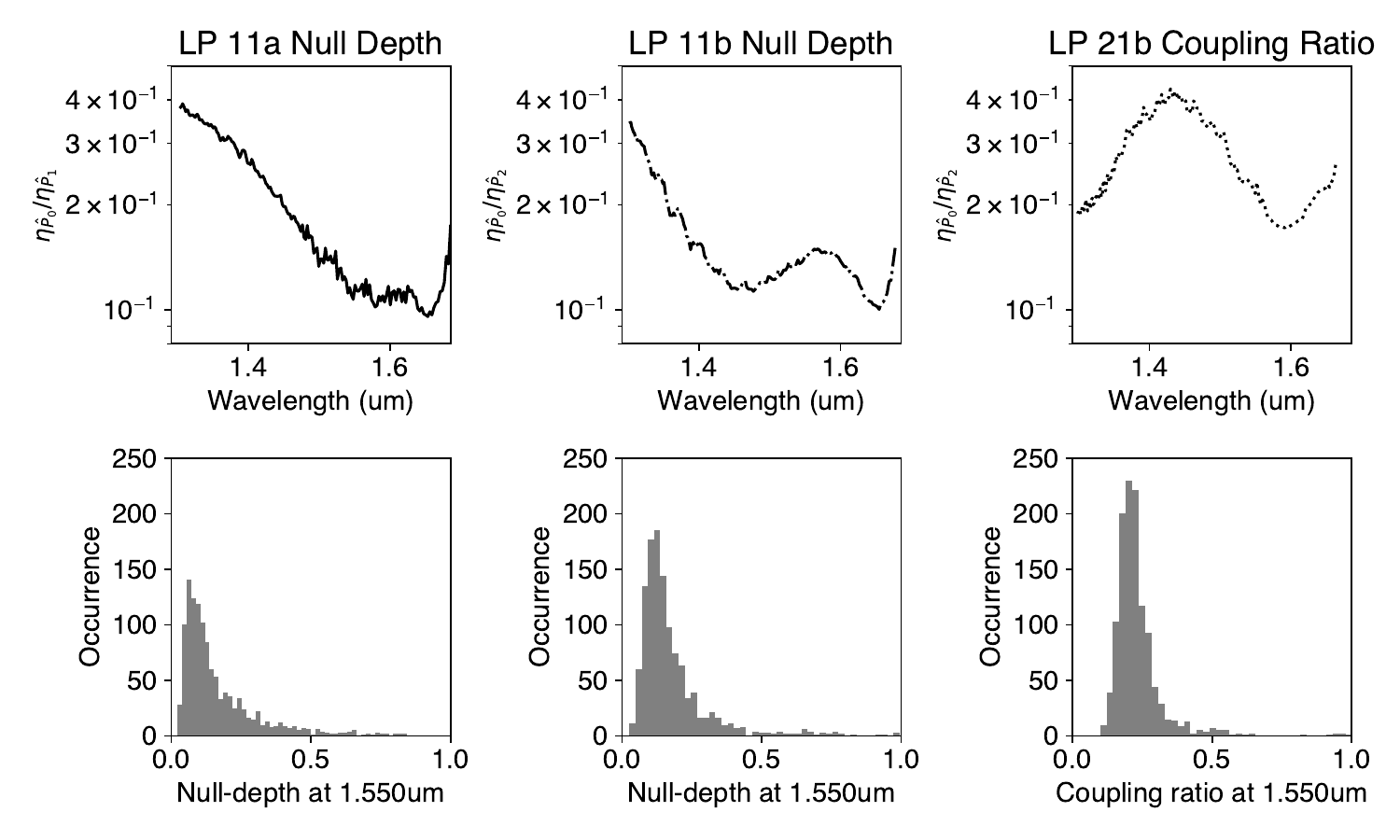}
	\caption{\label{fig:sky_nulldepths} Top) On-sky null depths measured for the LP11a (left) and LP11b (center) ports. The right panel shows the ratio of the LP21b coupling measured at $\hat{P}_0$ to that measured at $\hat{P}_2$. While not a formal null depth, it is a useful metric that characterizes the sensitivity of the LP21b mode to flux originating from $\hat{P}_2$. Bottom) Histograms of the corresponding frame-by-frame null depths or coupling ratios evaluated at the lantern design wavelength of $\lambda = 1.550~\mu$m. The pronounced asymmetry of the distributions suggests that the measured nulls are dominated by rapidly varying atmospheric and instrumental fluctuations. Such distributions can be analyzed with the null self-calibration technique, which can improve sensitivity to underlying astrophysical signals.
}
\end{center}
\end{figure*}

The on-sky null depths for the LP11a and LP11b ports are derived by taking the ratio of the median stellar coupling measured at $\hat{P}_0$, denoted $\eta_{\hat{P}_0}$, to the median peak coupling measured at $\hat{P}_1$ and $\hat{P}_2$, respectively. The resulting wavelength-dependent null depths are shown in Fig.~\ref{fig:sky_nulldepths}. For comparison, we also compute the ratio of the LP21b flux at $\hat{P}_0$ to its flux measured at $\hat{P}_2$. Although this quantity does not constitute a formal null depth, since $\hat{P}_2$ does not correspond to the location of maximum coupling for the LP21b mode, Fig.~\ref{fig:sky_data} demonstrates that this port receives substantial flux from a source located at $\hat{P}_2$. As a result, the ratio still provides useful information regarding the response of the instrument to an off-axis object.

An alternative approach is to compute frame-by-frame ratios by dividing each individual measurement of $\eta_{\hat{P}_0}$ by the corresponding median peak coupling value, rather than taking the median of the stellar-coupling measurements beforehand. Histograms of these distributions at the lantern design wavelength of 1.550~$\mu$m are presented in the bottom row of Fig.~\ref{fig:sky_nulldepths}. Future work could explore the application of the null self-calibration (NSC) technique to these distributions. By fitting a statistical model to the observed null-depth distribution, NSC can disentangle the contributions from the underlying astrophysical signal and from instrumental or atmospheric fluctuations, potentially improving measurement sensitivity \cite{hanot_improving_2011, martinod_scalable_2021}.

\section{CONCLUSION}

This work presents the first on-sky demonstration of the Photonic Lantern Nuller installed on the SCExAO platform. In this configuration, an AO-corrected beam is injected into the PLN, and the output modes are dispersed by a spectrograph operating at a resolving power of approximately $R \approx 160$. We describe the daytime calibration procedure and evaluate the instrument performance through laboratory measurements. We also present results from on-sky engineering observations, including the discovery and subsequent mitigation of a substantial alignment offset between the instrument as calibrated during the day and the actual on-sky observation. Following this realignment, spectra were acquired at three key locations: the central null and the positions of peak coupling for the LP11a and LP11b modes. Analysis of these measurements yields on-sky null depths for the LP11 ports on the order of $10^{-1}$. Additional observations were obtained in May 2025 on several known close-separation, low-contrast binary systems, together with unresolved reference stars for PSF calibration. Further analysis of these datasets is currently underway.

Future engineering efforts will focus on identifying the mechanisms responsible for alignment shifts between laboratory calibration and on-sky operation, as well as developing more robust procedures for optimally aligning the lantern for observations. Looking ahead, coupling the PLN to a higher-resolution spectrograph ($R \gg 4000$) would enable the application of high-resolution spectroscopic techniques that exploit resolved atomic and molecular absorption features. Such approaches have the potential to improve companion-detection sensitivity by approximately two orders of magnitude in flux ratio while simultaneously providing access to atmospheric characterization through the identification of specific molecular and atomic species.

\acknowledgments 
Based on data collected at Subaru Telescope, which is operated by the National Astronomical Observatory of Japan. The authors wish to recognize and acknowledge the very significant cultural role and reverence that the summit of Maunakea has always had within the indigenous Hawaiian community, and are most fortunate to have the opportunity to conduct observations from this mountain.

The development of SCExAO is supported by the Japan Society for the Promotion of Science (Grant-in-Aid for Research \#23340051, \#26220704, \#23103002, \#19H00703, \#19H00695 and \#21H04998), the Subaru Telescope, the National Astronomical Observatory of Japan, the Astrobiology Center of the National Institutes of Natural Sciences, Japan, the Mt Cuba Foundation and the Heising-Simons Foundation. The development of the CACAO software is supported by the National Science Foundation under award 2410616.

The development of AO3k is supported by the Japan Society for the Promotion of Science (Grant-in-Aid for Research \#19H00695 and \#21H04998), the Subaru Telescope, the National Astronomical Observatory of Japan and the Astrobiology Center of the National Institutes of Natural Sciences, Japan.

Y.X. acknowledges support from the National Science Foundation Graduate Research Fellowship under Grant No. 1122374. Additional effort has been supported by the National Science Foundation under Grant Nos. 2109231,  2109232, 2308360, and 2308361, and the Nederlandse Organisatie voor Wetenschappelijk Onderzoek Award No. 184.036.004.

Portions of this manuscript were edited with the assistance of ChatGPT (OpenAI), including the paraphrasing of text derived from the author's previous publications and writings. All scientific content, analysis, interpretations, and conclusions were developed and verified by the author.

\bibliography{report} 
\bibliographystyle{spiebib} 

\end{document}